\documentclass[useAMS]{emulateapj}

\usepackage{epsfig}
\usepackage{graphicx}
\usepackage{epstopdf}
\usepackage{amsmath}
\usepackage{amssymb}
\usepackage{appendix}

\def\be{\begin{equation}}
\def\ee{\end{equation}}
\def\ba{\begin{eqnarray}}
\def\ea{\end{eqnarray}}
\def\go{\mathrel{\raise.3ex\hbox{$>$}\mkern-14mu
             \lower0.6ex\hbox{$\sim$}}}
\def\lo{\mathrel{\raise.3ex\hbox{$<$}\mkern-14mu
             \lower0.6ex\hbox{$\sim$}}}

\begin{document}

\title{Magnetar Giant Flares and Their Precursors --- \\
Flux Rope Eruptions with Current Sheets
}

\author{Cong Yu\altaffilmark{1,3} and Lei Huang\altaffilmark{2,4}}
\altaffiltext{1}{National Astronomical Observatories/Yunnan
Astronomical Observatory, Chinese Academy of Sciences, Kunming,
650011, China; {\tt cyu@ynao.ac.cn}}
\altaffiltext{2}{Key
Laboratory for Research in Galaxies and Cosmology, Shanghai
Astronomical Observatory, Chinese Academy of Sciences, Shanghai,
200030, China; {\tt muduri@shao.ac.cn}}
\altaffiltext{3}{Key
Laboratory for the Structure and Evolution of Celestial Object,
Chinese Academy of Sciences, Kunming, 650011, China; }
\altaffiltext{4}{Key Laboratory of Radio Astronomy, Chinese
Academy of Sciences, China. }

\begin{abstract}
We propose a catastrophic magnetospheric model for magnetar
precursors and their successive giant flares. Axisymmetric models
of the magnetosphere, which contain both a helically twisted flux
rope and a current sheet, are established based on force-free
field configurations. In this model, the helically twisted flux
rope would lose its equilibrium and erupt abruptly in response to
the slow and quasi-static variations at the ultra-strongly
magnetized neutron star's surface. In a previous model without
current sheets, only one critical point exists in the flux rope
equilibrium curve. New features show up in the equilibrium curves
for the flux rope when current sheets appear in the magnetosphere.
The causal connection between the precursor and the giant flare,
as well as the temporary  re-entry of the quiescent state between
the precursor and the giant flare, can be naturally explained.
Magnetic energy would be released during the catastrophic state
transitions. The detailed energetics of the model are also
discussed. The current sheet created by the catastrophic loss of
equilibrium of the flux rope provides an ideal place for magnetic
reconnection. We point out the importance of magnetic reconnection
for further enhancement of the energy release during eruptions.

\end{abstract}


\section{Introduction}

Two closely related types of high-energy sources $-$ Anomalous
X-ray Pulsars (AXPs) and Soft Gamma-ray Repeaters (SGRs), are well
explained as magnetars, neutron stars with super-strong
($10^{14}-10^{15}$G) magnetic fields \citep{Maze79,DT92,Kouv98}.
It is commonly accepted that dissipation of the magnetic fields
drives persistent and bursting emission \citep{MS95,HK98,
TLK02,GKW02}. More rarely and unpredictably, more violent
outbursts $-$ giant flares, have been identified. Typically, a
giant flare releases a total energy of $\sim10^{44}-10^{46}$ ergs
and shows a peak luminosity over a million times the Eddington
luminosity of a neutron star \citep{WT06,Mere08}.

There exist two principal scenarios for the place where the
magnetic energy is accumulated before an eruptive outburst: in the
crust \citep{TD01} or in the magnetosphere \citep{Lyut06}. The
latter possibility was put forward to interpret the short
timescale of giant flare rise time, $\sim0.25\mathrm{ms}$
\citep{Palm05}. It has a distinguishing feature that the energy is
accumulated quasi-statically in the magnetosphere prior to the
eruption, on a timescale much longer than the dynamical timescale
of giant flares. The stored magnetic energy is limited by the
total external magnetic energy, instead of the tensile strength of
the crust \citep{Yu11b}. However, the origin of the catastrophic
state transitions, i.e., the transitions from a quasi-static
evolution (e.g. caused by flux injections and/or crust motions) to
a fast dynamical evolution, still remains a question for the
magnetospheric model. Recently, \citet{Yu12} investigated the
catastrophic mechansim of the helically twisted flux rope
eruption. The author showed that, with the gradual variation at
magnetar surface, either flux injections or crust gradual motions,
the flux rope will evolve correspondingly quasi-statically. Once
the flux rope reaches a critical height, it will become
dynamically unstable and erupt to produce the giant flare. The
catastrophic behavior naturally answers the above question.

Precursors to giant flares have been identified. For instance, the
2004 giant flare was proceeded by a 1 second long energetic
outburst event. The energy from the precursor is estimated to be
about $10^{42-43}$ ergs. After the precursor, the magnetar entered
a temporary quiescent state and stayed in the quiescent state for
about 140 seconds. Then it finally gave rise to the more violent
flare. It is inferred that the precursor and the giant flare are
causally related \citep{Hurl05}.
This precursor is hard to explain by our previous model,
since the catastrophic state transition could take place only once
due to the single critical point appearing in the equilibrium curve.
Additional physical elements should be included.

It is conceivable that, once the flux rope loses its equilibrium,
a current sheet can be generated in the magnetosphere. This type
of current sheet in magnetar magnetosphere has been hypothesized
by \citet{Lyut06,GH10}, which provides an ideal place for magnetic
reconnection. The magnetic reconnection is of vital importance to
the magnetic field dissipations \citep{PF00,GH10,MU12}, which
plays a crucial role not only in magnetars, but also in
rotation-driven pulsars in general, especially in the recent
observed crab nebula flares \citep[e.g.][]{SA12}. It will
contribute a sizable fraction of the total magnetic energy
dissipation \citep{Lyub96}. Secondary plasmoid instability is
expected to occur in the current sheet \citep{HB12}. The bursty,
non-steady character of the reconnection process marked by
plasmoid ejection \citep{Yu11a}  is likely to induce fast
variabilities in the magnetospheres. Unfortunately, no solid
calculations about magnetar magnetospheres with such current
sheets have been performed yet due to the complexity of this mixed
boundary value problem (the magnetar surface plus the current
sheet).

In this Letter, we try to explain the successive appearance of
precursors and giant flares with a magnetospheric model
incorporating a current sheet. This Letter is structured as
follows: we briefly describe in Section 2 the basic force-free
field configurations which contain both a flux rope and a current
sheet, as well as the equilibrium constraints. The catastrophic
responses of the flux rope to the magnetar surface variations are
described in Section 3. The magnetic energy release during
catastrophic state transitions is estimated in Section 4.
Conclusions are given in Section 5.

\section{Force-Free Magnetosphere with Both Current Sheet and Flux Rope}

The magnetosphere is assumed to be in a force-free (i.e.,
$\mathbf{J}\times\mathbf{B}=0$) equilibrium state \citep{Yu11a}.
The model is essentially the same as that in \citet{Yu12}, except
that we consider an additional current sheet in the magnetosphere.

\subsection{Basic Magnetic Configurations}

\begin{figure}
\includegraphics[scale=0.55]{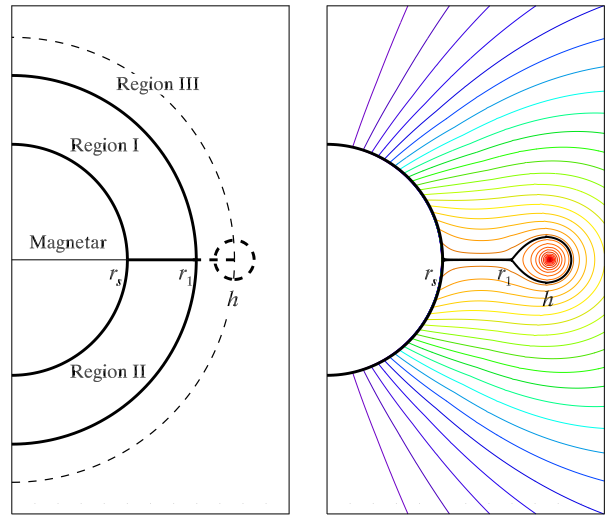}
\caption{\label{fig1} Magnetosphere containing a twisted flux rope
and a current sheet. {\it Left}: The radius of magnetar is denoted
by $r_s$. The current sheet is designated as the thick horizontal
line at the equator, the height of which is denoted by $r_1$. The
twisted flux rope is designated as thick dashed circle at the
equator, the height of which is denoted by $h$. The computation
domain is divided into three regions. Solutions in different
regions are obtained separately and finally melted together to
form a global solution. {\it Right}:  The magnetic field lines
(thin solid line) and the current sheet (thick solid line) are
shown. }
\end{figure}

We show the schematic diagram of our model in the left panel of
Fig.\ref{fig1}. The toroidal force-free magnetic flux rope, shown
by a thick dashed circle, is suspended by force balance in the
magnetosphere at the height, $h$, measured from the neutron star
center. In the interior of the helically twisted flux rope, the
force-free solution developed by \citet{Lund50} is used. A simple
relation between $r_0$ (minor radius of the rope) and $I$ (current
carried by the rope), i.e., $r_0=(r_{00}I_0)/I=r_{00}/J$, is valid
for such flux ropes. The quantity $I_0$ is related to the magnetar
radius $r_s$ and a constant $\Psi_0$ (with magnetic flux
dimension) by $I_0=(\Psi_0c)/r_s$. Here $c$ stands for the speed
of light. The dimensionless current $J=I/I_0$ is the current
measured in unit of $I_0$. The parameter $r_{00}$ is fixed as
0.01\footnote{For convenience of numerical calculations, lengths
are measured in $r_s$, magnetar radius. Currents are measured in
$I_0$ and the magnetic fluxes in $\Psi_0$.}, which is the value of
$r_0$ for $J=1$.

The stationary axisymmetric field configurations outside the flux
rope can be expressed in terms of the magnetic stream function
$\Psi(r,\theta)$ as ${\bf B}=\nabla\Psi\times\nabla\phi$. Note
that we adopt spherical polar coordinates $(r,\theta,\phi)$ in our
calculations. The Grad-Shafranov (GS) equation can be derived
according to the force-free constraint. Explicitly, it reads
\begin{equation}\label{inhomoGS}
\frac{\partial^2\Psi}{\partial r^2}
+\frac{\sin\theta}{r^2}\frac{\partial}{\partial\theta}
\left(\frac{1}{\sin\theta}\frac{\partial\Psi}{\partial\theta}\right)
=-(r\sin\theta)\frac{4\pi}{c}J_{\phi}\ ,
\end{equation}
where $J_{\phi}$ is the current density caused by the flux rope.
We treat it as a circular ring current of the following form,
$J_{\phi}=(I/h)\delta(\cos\theta)\delta(r-h)$ \citep{Yu12}. For
simplicity, we use a dipolar boundary condition at the magnetar
surface, i.e., $\Psi(r_s,\theta) =
\Psi_0\sigma\left(1-\cos^2\theta\right)$, where $\sigma$ is a
dimensionless quantity which dictates the magnitude of the flux at
the magnetar surface.

Besides the twisted flux rope, we also include an equatorial
current sheet in our model. To account for the presence of current
sheet, a second boundary condition needs to be satisfied,
$\Psi(r,\pi/2)=\Psi_0\sigma$ for $r_s\leq r\leq r_1$, where $r_1$
denotes the tip of the current sheet. In the left panel of
Fig.\ref{fig1}, the horizontal thick solid line at the equator
between $r_s$ and $r_1$ represents the current sheet. Solutions to
the GS equation in spherical coordinates become non-trivial due to
the second boundary condition \citep{Lin98}. Our numerical
strategy is domain decomposition: the GS equation is solved in
three different regions in the computational domain, labelled as
region I, II, and III. The regions I and II are located between
two thick solid semi-circles, $r=r_s$ and $r=r_1$, and separated
by the current sheet. The flux rope lies in the region III with
$r>r_1$. Solutions in the three different regions are matched
together to form a global solution. An illustrative example about
our basic magnetic configuration is shown in the right panel of
Fig.\ref{fig1}. Both field lines (thin solid line) and the current
sheet (thick solid line) are shown. The Y-point condition,
$B_\theta(r_1,\pi/2) = 0$, must be satisfied at the tip of the
current $(r_1,\pi/2)$.

\subsection{Equilibrium Constraints}
In our catastrophic eruption model, the flux rope's dynamical
behavior on short timescale is triggered by the slow and
quasi-static evolution at the neutron star surface. Before the
catastrophic state transition occurs, the flux rope is assumed to
be in a quasi-static equilibrium. In the following, we give a
brief description of the two aspects of the equilibrium
constraint.

The first one is the force-balance condition, which is fulfilled
if the forces exerted on the flux rope cancel each other. Simply
put, the magnetic field generated by the ring current within the
flux rope, $B_{\mathrm{s}}$=$\frac{I}{ch}(\ln\frac{8 h}{r_0}-1)$
\citep{Shaf66}, must be balanced by the external magnetic field
$B_\mathrm{e}$ at $(r=h,\theta=\pi/2)$. The requirement that
$B_{\mathrm{e}}-B_{\mathrm{s}}=0$ can be expressed as an equation
$f(\sigma,J,h)=0$ (For details, please refer to Yu (2012)).

The second one involves the ideal frozen-flux condition, which
connects the current within the flux rope and the boundary
conditions at the neutron star surface. It demands that the
variable $\Psi(r,\theta)$ on the edge of the flux rope $r=h-r_0$
at the equator keeps constant  in the course of the flux rope's
quasi-static evolution. We define another function $g(\sigma,J,h)$
to represent the value of $\Psi(h-r_0,\pi/2)$. Then the
frozen-flux requirement can be explicitly expressed as
$g(\sigma,J,h)=\mathrm{cosnt}$.

Combine the two aspects of the equilibrium constraint and we
arrive at
\begin{equation}\label{eqm}
\left\{\begin{array}{lll}
f(\sigma,J,h)=0\\
g(\sigma,J,h)=\mathrm{const}. \\
\end{array}\right.\
\end{equation}
The equilibrium curve in the following section can be obtained by
solving this equation numerically \citep{Yu12}.

\section{Catastrophic Response of Flux Ropes to Changes at Magnetar Surface}
Slowly progressing events which occur at the neutron star surface,
either the crust horizontal motion \citep{Rude91,TD01,Jone03}, or
new magnetic fluxes continuously injected into the magnetosphere
\citep{KR98,TLK02,Lyut06,GMH07}, would trigger the catastrophic
outburst of the flux rope. In this Letter, we focus on one aspect
for brevity, i.e., the flux injections. The background magnetic
field would decrease (increase) gradually, if new current-carrying
magnetic fluxes of the opposite (same) polarity are injected. Two
kinds of magnetic configurations in the magnetosphere, inverse and
normal, are found in \citet{Yu12}, depending on the polarity of
the neutrons star's surface magnetic flux. In the normal
configuration the equilibrium position of the flux rope is too
close to the magnetar surface and the regular variability at the
magnetar surface would contaminate the behavior of the flux rope.
As a result, we will focus in this Letter on the inverse
configurations.

\begin{figure}
\includegraphics[scale=0.85]{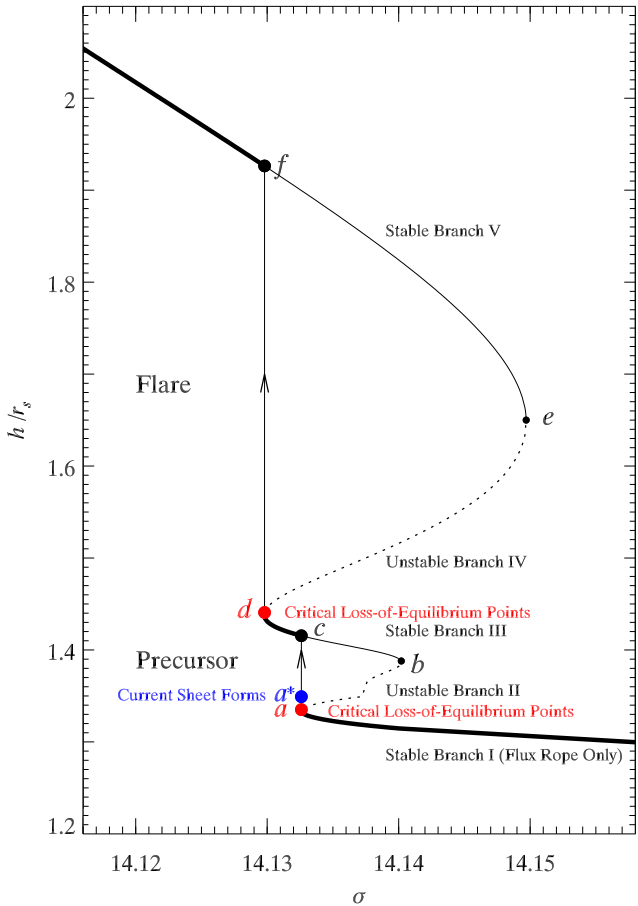}
\caption{\label{fig2} Equilibrium curve for a flux rope in
magnetosphere with a current sheet. The two red dots $a$ and $d$
represent the critical loss-of-equilibrium points. Only the lower
one $a$ remains if the current sheet is not included. The first
jump of the twisted flux rope occurs from $a$ to $c$, where the
flux rope reaches a lower stable branch, with the current sheet
forming at the location marked with the blue point $a^*$. The
second jump occurs from $d$ to $f$, where the flux rope and
current sheet reaches an upper stable branch. The two jumps are
thought to correspond to the precursor and the flare,
respectively. }
\end{figure}

We investigate in detail the responses of the flux rope's
equilibrium height $h$ to the changes of $\sigma$ ($\sigma$
reflects the background field magnitude) for the inverse magnetic
configurations. Numerical results of Equation (\ref{eqm}) are
shown in a curve with thick solid segments and thin dotted
segments in Fig.\ref{fig2}. Five branches alternatively appear
from bottom to top, named from I to V respectively, in the
equilibrium curve according to the curve's slope in the
$\sigma$-$h$ plane. The magnetic force analysis shows that the
branches with negative slope in the $\sigma$-$h$ diagram ($h$
increases with decreasing $\sigma$) are stable, while the branches
with positive slope are unstable \citep{Yu12}. There are three
stable branches, I, III, and V, and two unstable branches. II and
IV, in this equilibrium curve. When the surface magnetic field at
the equator satisfies $B_{\theta}(r_s,\pi/2)=0$, the current sheet
begins to form in the magnetosphere (see blue point $a^*$ in
Fig.\ref{fig2}). When the height of the flux rope is lower than
this point, i.e., prior to the current sheet formation, the
magnetic field configurations are specified in the same manner as
\citet{Yu12}. Higher than this point, the magnetic field
configurations are determined according to the the method outlined
in Section 2. Additional unstable branches II (between points
'$a$' and '$b$'), stable branches III (between points '$b$' and
'$d$'), unstable branches IV (between points '$d$' and '$e$'), and
stable branches V (higher than point '$e$') show up in the
equilibrium curve.

The thick solid segments in Fig.\ref{fig2} on the stable branches
show the $\sigma$-$h$ evolution of the flux rope. Compared to our
previous work without current sheets, new features appear in the
properties of the equilibrium curve. The biggest difference is
that there appear two critical points, '$a$' and '$d$' (denoted by
red dots\footnote{Critical points separate the stable and unstable
branches of the equilibrium curve and the instability threshold
lies at the critical points.}), in the equilibrium curve. As
$\sigma$ decreases, the flux rope moves leftward and upward along
the equilibrium curve until it reaches the first critical point
'$a$' at $\sigma=14.1326,h/r_s=1.335$. The stability of the
equilibrium in this system breaks after this point. During this
'loss-of-equilibrium' process $\sigma$ can be regarded as
unchanged because this would occur on a dynamical timescale and
$\sigma$ varies on a much longer quasi-static timescale. As a
result, the flux rope would jump to another stable branch III with
a larger equilibrium height in the equilibrium curve (see black
point '$c$'). On this stable branch, the flux rope can evolve on a
timescale much longer than the dynamical timescale. With the
further injection of magnetic flux from below, i.e., the decrease
of $\sigma$, the flux rope would stay on this new stable branch
for some time. When the flux rope gradually approaches the second
critical point '$d$' at $\sigma=14.1298,h/r_s=1.435$, the flux
rope could not maintain its stable equilibrium and would jump to
stable branch V (see black point '$f$'). The two vertical jumps
from critical points to another stable branches represent the
catastrophic state transitions. The first state transition from
'$a$' to '$c$' and the second one from '$d$' to '$f$' correspond
to precursor and giant flare, respectively. The temporary
quiescent state between the precursor and the giant flare
corresponds to the thick solid segment on stable branch III after
the first state transition. As a result, the causal connection
between the precursor and the giant flare can be naturally
explained by our model. When the height of the flux rope becomes
larger, we find that the stable branch V asymptotically approaches
a vertical line in the $\sigma$-$h$ plane. The value of $\sigma$
of this vertical line can be regarded as a threshold. If the
surface magnetic flux is less than this threshold, the flux rope
would reach infinity quasi-statically. Such ideal behavior occurs
only when magnetic reconnection is strictly prohibited. However,
this quasi-static behavior would be replaced by a dynamical one if
magnetic reconnection proceeds sufficiently fast.

\section{Energy Release During Catastrophic State Transitions}

The catastrophic behavior of the flux rope also induces energy
release during the state transitions. It is instructive to know
the energetics of the catastrophic model. Observationally, the
magnetic energy of magnetars is about ${\rm E}_{\rm mag}\sim
10^{48}({\rm B}/{10^{15}{\rm G}})^2({\rm R}/10{\rm km})^3$ ergs. The
biggest giant flare ever observed is from SGR1806-20, the energy
from which is $\sim10^{46}$ ergs. So only $1\%$ of the
magnetic energy release is already enough to explain the giant flares.
A precursor with energy of $\sim10^{42-43}$ ergs was also observed
preceded this giant flare. The energy release
fraction compared to the total energy is about $10^{-6}\sim10^{-5}$.

The total magnetic energy of the magnetosphere, $W_t$, is
estimated as follows,
\begin{equation}
W_t(h)=-\int_{h}^{\infty}F(h^{\prime})dh^{\prime}+
W_\mathrm{dipole} \ ,
\end{equation}
where $F=2\pi Ih(B_\mathrm{s}-B_\mathrm{e})/c$ is the total force
exerted on the flux rope and $W_\mathrm{dipole}$ is the magnetic
energy of the background dipolar field. Note that in the above
equation, we perform the work integration along the path where
$\sigma$ holds constant.

The energy release fraction during the ideal loss of equilibrium
due to the first jump, $[W_t(h_a)-W_t(h_c)]/W_t(h_a)$ is about
$4\times10^{-5}$ (subscripts $a$ and $c$ correspond to points $a$
and $c$ in Fig.\ref{fig2}). For a magnetar with magnetic field
$B\sim 10^{15}$ Gauss, the energy release of the precursor in our
model is about $4\times10^{43}$ erg, which is already enough to
explain the precursor energy release inferred from observations
\citep{Hurl05}. After the precursor, the flux rope lies on a
stable branch with larger height of the equilibrium curve again.
On this branch, the flux rope transits to a slow evolution stage
again. This is consistent with observations that the magnetar
re-enter a temporary quiescent state. With the further magnetic
flux injection, the flux rope height $h$ further increases and
reach the second critical point. At this point, the second jump
occurs, the energy release fraction of which,
$[W_t(h_d)-W_t(h_f)]/W_t(h_d)$ (subscripts $d$ and $f$ correspond
to points $d$ and $f$ in Fig.\ref{fig2}) is higher than the first
jump, approximately $0.05\%$. Though this value is still below the
required $1\%$ level to explain the giant flares, it is worthwhile
to note that in our calculation the energy release is completely
ideal. Most of the energy is locked in current sheet and we do not
take into account the non-ideal effects of magnetic reconnection.
When we consider the effects of magnetic reconnection, the
magnetic energy lock in the current sheet would be released, and
the energy release fraction can be further enhanced. This is left
for a future study.

\section{Conclusion}

In this Letter we focus on the possibility of magnetospheric
origin for the precursors and the successive giant flares. We put
forward a force-free magnetosphere model containing a helical flux
rope and a current sheet below the flux rope. The catastrophic
response of the flux rope is examined in detail, taking into
account the gradual process of flux injections at the
ultra-strongly magnetized neutron star surface. In this model, the
twisted flux rope would lose its equilibrium due to the
quasi-static evolutions at the surface and finally erupt abruptly.
We find that there may exist two critical points in the flux rope
equilibrium curve when current sheets are taken into account.
According to this new feature, the precursor and the giant flares
can be naturally explained as two stages in the evolution of our
models (see Section 3). The dynamical state transitions around the
two critical points correspond to the precursor and the giant
flares, respectively. The stable branch between the two
transitions represents the quiescent state between the precursor
and the giant flare.

The detailed energetics of the model are also discussed. The
magnetic energy release fraction in the first jump is consistent
with the precursor energy budget. The energy release fraction in
the second jump, which corresponds to the giant flare is lower
than the value inferred from observations. This shows that
additional energy release is necessary to account for the giant
flare. The current sheet generated by the catastrophic loss of
equilibrium behavior of the flux rope also provides an ideal place
for magnetic reconnection, which can further enhance the energy
release during the eruptions. How magnetic reconnection affects
the energy release needs further investigations. Another
possibility to enhance the energy release fraction depends on the
boundary condition at the magnetar surface. Prior results show
that dipolar boundary condition is the least efficient to release
magnetic energy \citep{FPI94}. How do boundary conditions, such as
the multipolar boundary conditions used in \citet{Yu12}, affect
the magnetic energy release is worth a further study.

The time-dependent numerical simulation of the our model is
another interesting issue \citep{Yu11a}. Based on time-dependent
models, the detailed light curves could be obtained from dynamical
simulations and be compared with observations. Thus certain model
parameters can be further constrained by the observational light
curves of giant flares.

\acknowledgments We would like to thank the anonymous referee for
instructive comments and helpful suggestions. This work has been
supported by National Natural Science Foundation of China (Grants
10703012, 11173057, 11033008, 11203055, 10625314, 11121062, and
11173046), Yunnan Natural Science Foundation (Grant 2012FB187) and
Western Light Young Scholar Programme of CAS and partly supported
by China Ministry of Science and Technology under State Key
Development Program for Basic Research (2012CB821800), the
CAS/SAFEA International Partnership Program for Creative Research
Teams, and the Strategic Priority Research Program on Space
Science, the Chinese Academy of Sciences (Grant No. XDA04060700).
Part of the computation is performed at HPC Center, Kunming
Institute of Botany, CAS, China.




\begin{thebibliography}{99}

    \bibitem[\protect\citeauthoryear{Duncan \& Thompson}{1992}]{DT92}
        Duncan, R. C., \& Thompson, C., 1992, ApJL, 392, 9

    \bibitem[\protect\citeauthoryear{Forbes, Priest, \& Isenberg}{Forbes et al.}{1994}]{FPI94}
        Forbes, T. G., Priest, E. R., \& Isenberg, P. A. 1994, Sol. Phys., 150, 245

    \bibitem[\protect\citeauthoryear{Gavriil, Kaspi, \& Woods}{Gavriil et al.}{2002}]{GKW02}
        Gavriil, F. P., Kaspi, V. M., Woods, P. M. 2002, Nature, 419, 142

    \bibitem[\protect\citeauthoryear{Gill \& Heyl}{2010}]{GH10}
        Gill, R., \& Heyl, J. S. 2010, MNRAS, 407, 1926

    \bibitem[\protect\citeauthoryear{G\"{o}tz,  Mereghetti, \& Hurley}{G\"{o}tz et al.}{2007}]{GMH07}
        G\"{o}tz, D.,  Mereghetti, S., \& Hurley, K. 2007, Ap\&SS, 308, 51

    \bibitem[\protect\citeauthoryear{Heyl \& Kulkarni}{1998}]{HK98}
        Heyl, J. S., \& Kulkarni, S. R. 1998, ApJL, 506, 61

    \bibitem[\protect\citeauthoryear{Huang \& Bhattacharjee}{2012}]{HB12}
        Huang Y. M., \& Bhattacharjee A., 2012, Phy. Rev. Lett., 109, 265002

    \bibitem[\protect\citeauthoryear{Hurley et al.}{2005}]{Hurl05}
        Hurley, F., et al. 2005, Nature, 434, 1098

    \bibitem[\protect\citeauthoryear{Jones}{2003}]{Jone03}
        Jones, P. B. 2003, ApJ, 595, 342

    \bibitem[\protect\citeauthoryear{Klu\'{z}niak \& Ruderman}{1998}]{KR98}
        Klu\'{z}niak, W. \& Ruderman, M., 1998, ApJL, 505, 113

    \bibitem[\protect\citeauthoryear{Kouveliotou et al.}{1998}]{Kouv98}
        Kouveliotou, C., et al. 1998, Nature, 393, 235

    \bibitem[\protect\citeauthoryear{Lin et al.}{1998}]{Lin98}
        Lin, J., Forbes, T. G., Isenberg, P. A., \& D\'{e}moulin, P. 1998, ApJ, 504, 1006

    \bibitem[\protect\citeauthoryear{Lundquist}{1950}]{Lund50}
        Lundquist, S. 1950, Ark. Fys., 2, 361

    \bibitem[\protect\citeauthoryear{Lyubarsky}{1996}]{Lyub96}
        Lyubarsky, Y. E. 1996, A\&A, 311, 172

    \bibitem[\protect\citeauthoryear{Lyutikov}{2006}]{Lyut06}
        Lyutikov, M. 2006, MNRAS, 367, 1602

    \bibitem[\protect\citeauthoryear{Mazets et al.}{1979}]{Maze79}
        Mazets, E. P., et al. 1979, Nature, 282, 587

    \bibitem[\protect\citeauthoryear{McKinney \& Uzdensky}{2012}]{MU12}
        McKinney, J. C. \& Uzdensky, D. A. 2012, MNRAS, 419, 573

    \bibitem[\protect\citeauthoryear{Mereghetti}{2008}]{Mere08}
        Mereghetti, S. 2008, A\&AR, 15, 225

    \bibitem[\protect\citeauthoryear{Mereghetti \&  Stella}{1995}]{MS95}
        Mereghetti, S. \&  Stella, L. 1995, ApJL, 442, 17

    \bibitem[\protect\citeauthoryear{Palmer et al.}{2005}]{Palm05}
        Palmer, D. M., et al. 2005, Nature, 434, 1107

    \bibitem[\protect\citeauthoryear{Priest \& Forbes}{2000}]{PF00}
        Priest E., \& Forbes T. 2000, Magnetic Reconnection. MHD Theory and Applications. Cambridge Univ. Press, Cambridge

    \bibitem[\protect\citeauthoryear{Ruderman}{1991}]{Rude91}
        Ruderman, M. 1991, ApJ, 366, 261

    \bibitem[\protect\citeauthoryear{Shafranov}{1966}]{Shaf66}
        Shafranov, V. D. 1966, Rev. Plasma Phys., 2, 103

    \bibitem[\protect\citeauthoryear{Sturrock \& Aschwanden}{2012}]{SA12}
        Sturrock, P. \& Aschwanden, M. J. 2012, ApJL, 751, 32

    \bibitem[\protect\citeauthoryear{Thompson \& Duncan}{2001}]{TD01}
        Thompson, C., \& Duncan, R. C. 2001, ApJ, 561, 980

    \bibitem[\protect\citeauthoryear{Thompson, Lyutikov, \& Kulkarni}{Thompson et al.}{2002}]{TLK02}
        Thompson, C., Lyutikov, M., \& Kulkarni, S. R. 2002, ApJ, 574, 332

    \bibitem[\protect\citeauthoryear{Woods \& Thompson}{2006}]{WT06}
        Woods, P. M., \& Thompson, C., 2006, in Compact Stellar X-Ray Sources, ed. W. H. G. Lewin \& van der Klis (Cambridge Univ. Press), 547

    \bibitem[\protect\citeauthoryear{Yu}{2011a}]{Yu11a}
        Yu, C., 2011a, MNRAS, 411, 2461

    \bibitem[\protect\citeauthoryear{Yu}{2011b}]{Yu11b}
        Yu, C., 2011b, ApJ, 738, 75

    \bibitem[\protect\citeauthoryear{Yu}{2012}]{Yu12}
        Yu, C., 2012, ApJ, 757, 67

\end{thebibliography}
\end{document}